\documentclass[a4paper]{article}
\usepackage[margin=2cm]{geometry}
\usepackage{graphicx}
\usepackage{amsmath}

\begin{document}

\title{Music Generation by Deep Learning\\
-- Challenges and Directions\footnote{To appear in Special Issue on Deep learning for music and audio,
	Neural Computing \& Applications,
	Springer Nature,
	2018.}}

\author{Jean-Pierre Briot$^\dagger$
\and Fran\c{c}ois Pachet$^\ddagger$
}

\date{$^\dagger$ Sorbonne Universit\'es, UPMC Univ Paris 06, CNRS, LIP6, Paris, France\\
{\tt Jean-Pierre.Briot@lip6.fr}\\[.2cm]
$^\ddagger$ Spotify Creator Technology Research Lab, Paris, France\\
{\tt francois@spotify.com}
}

\maketitle

{\bf Abstract:}
In addition to traditional tasks such as prediction, classification and translation, deep learning is receiving growing attention as an approach
for music generation,
as witnessed by recent research groups such as Magenta at Google and CTRL (Creator Technology Research Lab) at Spotify.
The motivation is in using the capacity of deep learning architectures and training techniques to automatically learn musical styles from arbitrary
musical corpora and then to generate samples from the estimated
distribution.
However,
a direct application of deep learning
to generate content
rapidly reaches
limits as
the generated content
tends to mimic the
training set
without
exhibiting
true creativity.
Moreover, deep learning architectures do not offer direct ways for controlling generation
(e.g., imposing some tonality or other arbitrary constraints).
Furthermore, deep learning architectures alone are autistic automata which generate music autonomously without human user interaction,
far from the objective of interactively assisting musicians to compose and refine music.
Issues such as: control, structure, creativity and
interactivity
are the focus of our analysis.
In this paper, we select some limitations of a direct application of deep learning to music generation,
analyze why the issues are not fulfilled
and how to address them by possible approaches.
Various examples of recent systems are cited as examples of promising directions.

\section{Introduction}
\label{section:introduction}

\subsection{Deep Learning}

Deep learning\index{Deep learning} has become a fast growing domain
and is now used routinely for classification\index{Classification} and prediction\index{Prediction} tasks,
such as image and voice recognition, as well as translation.
It
emerged about 10 years ago,
when a deep learning architecture
significantly outperformed standard techniques using handcrafted features
on an image classification task \cite{hinton:fast:algorithm:2006}.
We may explain this success and reemergence of artificial neural networks\index{Neural network} architectures and techniques
by the combination of:

\begin{enumerate}

\item {\em technical progress}, such as:
convolutions\index{Convolution|see{Convolutional neural network}},
which provide motif translation invariance \cite{le:cun:convolutional:handbook:1998},
and LSTM (Long Short-Term Memory),
which resolved
inefficient training of recurrent neural networks \cite{hochreiter:lstm:1997};

\item availability of multiple {\em data sets};

\item
	availability of {\em efficient and cheap computing power},
	e.g., offered by graphics processing units (GPU).

\end{enumerate}

There is no consensual definition
of deep learning\index{Deep learning}.
It is a repertoire of machine learning\index{Machine learning} (ML\index{ML}) techniques,
based on artificial neural networks\index{Artificial neural network}\footnote{With many variants
	such as convolutional networks\index{Convolutional network},
	recurrent networks\index{Recurrent network},
	autoencoders\index{Autoencoder},
	restricted Boltzmann machines\index{Restricted Boltzmann machine}, etc. \cite{goodfellow:deep:learning:book:2016}.}.
The common ground is the term {\em deep},
which means that there are multiple layers\index{Layer} processing multiple
levels of abstractions\index{Abstraction},
which are automatically extracted from data,
as a way to express complex representations in terms of simpler representations.

Main applications of deep learning are within the two traditional machine learning tasks\index{Task} of
{\em classification}\index{Classification} and {\em prediction}\index{Prediction},
as a testimony of the initial DNA of neural networks:
logistic regression\index{Logistic regression} and linear regression\index{Linear regression}.
But a growing area of application of deep learning techniques is the {\em generation} of {\em content\index{Content}}:
text\index{Text}, images\index{Image}, and {\em music\index{Music}}, the focus of this article.

\subsection{Deep Learning for Music Generation}

The motivation for using deep learning, and more generally machine learning techniques, to generate musical content
is its generality.
As opposed to handcrafted models for, e.g., grammar-based
\cite{steedman:generative:grammar:blues:mp:1984}
or rule-based music generation systems
\cite{ebcioglu:expert:system:bach:cmj:1988},
a machine-learning-based generation system can automatically learn a model, a {\em style},
from an arbitrary corpus of music.
Generation can then take place by using prediction
(e.g., to predict the pitch of the next note of a melody)
or classification
(e.g., to recognize the chord corresponding to a melody),
based on the distribution and correlations learnt by the deep model which represent the style of the corpus.

As stated by Fiebrink and Caramiaux in \cite{fiebrink:ml:creative:tool:arxiv:2016},
benefits are:
1) it can make creation feasible when the desired application is too complex to be described by analytical formulations or
manual brute force design;
2) learning algorithms are often less brittle than manually-designed rule sets
and learned rules are more likely to generalize accurately to new contexts in which inputs may change.



\subsection{Challenges}
\label{section:challenges}

A direct application of deep learning
architectures and techniques
to generation,
although it could produce impressing results\footnote{Music difficult to
	distinguish from the original corpus.},
suffers from some limitations.
We consider here\footnote{Additional challenges are analyzed
	in \cite{briot:dlt4mg:springer:2018}.}:

\begin{itemize}

%


\item {\em Control}, e.g., tonality conformance, maximum number of repeated notes, rhythm, etc.;

\item {\em Structure}, versus wandering music without a sense of direction;

\item {\em Creativity}, versus imitation and risk of
plagiarism;


\item {\em Interactivity}, versus automated single-step generation.



\end{itemize}

\subsection{Related Work}
\label{section:related:work}

A comprehensive survey and analysis by Briot {\em et al.} of deep learning techniques to generate musical content
is available in a book
\cite{briot:dlt4mg:springer:2018}.
In \cite{herremans:taxonomy:music:generation:acm:cs:2017},
Herremans {\em et al.} propose a function-oriented taxonomy for various kinds of music generation systems.
Examples of surveys about of AI-based methods for algorithmic music composition are
by Papadopoulos and Wiggins \cite{papadopoulos:ai:algorithmic:composition:1999}
and by Fern\'andez and Vico \cite{fernandez:ai:methods:algorithmic:composition:survey:jair:2013},
as well as books by Cope \cite{cope:algorithmic:composer:book:2000} and by Nierhaus \cite{nierhaus:algorithmic:composition:book:2009}.
In \cite{graves:generating:sequences:rnn:arxiv:2014},
Graves analyses the application of recurrent neural networks architectures to generate sequences (text and music).
In \cite{fiebrink:ml:creative:tool:arxiv:2016}, Fiebrink and Caramiaux address the issue of using machine learning to generate creative music.
We are not aware of a comprehensive analysis dedicated to deep learning
(and artificial neural networks techniques)
that systematically analyzes limitations and challenges,
solutions and directions,
in other words that is {\em problem-oriented} and not just application-oriented.

\subsection{Organization}
\label{section:organization}

The article is organized as follows.
Section~\ref{section:introduction} (this section) introduces the general context of deep learning-based music generation
and lists some important challenges.
It also includes a comparison to some related work.
The following sections analyze each challenge and some solutions,
while illustrating through examples of actual systems:
control/section~\ref{section:control},
structure/section~\ref{section:structure},
creativity/section~\ref{section:creativity}
and interactivity/section~\ref{section:interactivity}.

\section{Control}
\label{section:control}

Musicians usually want to adapt ideas and patterns borrowed from other contexts
to their own objective,
e.g., transposition to another key, minimizing the number of notes.
In practice this means the ability to control generation by a deep learning architecture.


\subsection{Dimensions of control strategies}
\label{section:control:dimensions:strategies}

Such arbitrary control is actually a
difficult issue for current deep learning architectures and techniques,
because standard neural networks are not designed to be controlled.
As opposed to Markov
models
which have an operational model where one can attach constraints onto their internal operational structure
in order to control the generation\footnote{Two examples are Markov constraints
	\cite{pachet:markov:constraints:ijcai:2011}
	and factor graphs \cite{pachet:variations:structured:ismir:2017}.},
neural networks
do not offer such an operational entry point.
Moreover, the distributed nature of their representation does not provide a direct correspondence to the structure of the content generated.
As a result, strategies for controlling deep learning generation that we will analyze have to rely on some {\em external} intervention at various
{\em entry points\index{Entry point}} (hooks\index{Hook}), such as:

\begin{itemize}

\item {\em Input};

\item {\em Output};


\item {\em Encapsulation/reformulation}.

\end{itemize}

\subsection{Sampling}
\label{section:control:sampling}

{\em Sampling} a model\footnote{The model can be stochastic,
	such as a restricted Boltzmann machine (RBM)
	\cite{goodfellow:deep:learning:book:2016},
	or deterministic, such as a feedforward or a recurrent network.
	In that latter case, it is common practice to sample from the softmax output
	in order to introduce {\em variability} for the generated content \cite{briot:dlt4mg:springer:2018}.}
to generate content
may be an entry point for control if we introduce {\em constraints} on the output generation
(this is called {\em constraint sampling}).
This is usually implemented by a generate-and-test\index{Generate-and-test} approach,
where valid solutions are picked from a set of generated random\index{Random} samples\index{Sample} from the model\footnote{Note that
	this may be a very costly process and moreover with no guarantee to succeed.}.
As we will see, a key
issue is how to guide the sampling process in order to fulfill the objectives (constraints),
thus sampling will be often combined with other strategies.

\subsection{Conditioning}
\label{section:control:conditioning}


The strategy of {\em conditioning\index{Conditioning}} (sometimes also named {\em conditional architecture\index{Conditional architecture}})
is to condition\index{Condition} the architecture on some extra conditioning information,
which could be arbitrary, e.g., a class label\index{Label} or data from other modalities\index{Modality}.
Examples are:

\begin{itemize}

\item a {\em bass line\index{Bass!line}} or a {\em beat\index{Beat} structure},
in the rhythm generation system \cite{makris:rhythm:composition:2017};

\item a {\em chord progression\index{Chord!progression}},
in the MidiNet\index{MidiNet} architecture \cite{yang:midinet:ismir:2017};

\item a {\em musical genre\index{Genre}} or an {\em instrument\index{Instrument}},
in the WaveNet\index{WaveNet} architecture \cite{oord:wavenet:arxiv:2016};

\item a set of {\em positional constraints\index{Positional constraint}},
in the Anticipation-RNN\index{Anticipation-RNN} architecture \cite{hadjeres:anticipation:rnn:arxiv:2017}.

\end{itemize}


In practice, the conditioning information
is usually fed into the architecture as an additional input layer\index{Input!layer}.
%
Conditioning is a way to have some degree of {\em parameterized control} over the generation\index{Generation} process.

\subsubsection{Example 1: WaveNet Audio Speech and Music Generation}
\label{section:systems:wavenet}

The WaveNet\index{WaveNet} architecture by
van der Oord {\em et al.} \cite{oord:wavenet:arxiv:2016}
is aimed at generating raw audio\index{Audio} waveforms\index{Waveform}.
The architecture
is based on a convolutional\index{Convolutional!network}
feedforward network without pooling\index{Pooling} layer\footnote{An important specificity of the architecture
	(not discussed here)
	is the notion of {\em dilated convolution\index{Dilated convolution}},
	where convolution filters are incrementally dilated in order to provide very large receptive fields with just a few layers,
	while preserving input resolution and computational efficiency \cite{oord:wavenet:arxiv:2016}.}.
It has been experimented on generation for three audio domains:
multi-speaker\index{Speaker}, text-to-speech\index{Text-to-speech} {(TTS\index{TTS})}
and music.

The WaveNet architecture
uses conditioning\index{Conditioning}
as a way to guide the generation, by adding an additional tag\index{Tag}
as a conditioning input\index{Conditioning!input}.
Two options are considered:
{\em global} conditioning or {\em local} conditioning, depending if the conditioning input is shared for {\em all} time steps\index{Time!step}
or is specific to {\em each} time step.

An example of application of conditioning WaveNet for a text-to-speech\index{Text-to-speech} application domain
is to feed linguistic features
(e.g., North American English or Mandarin Chinese speakers)
in order to generate speech with a better prosody.
The authors
also
report preliminary experiments on conditioning\index{Conditioning} music models
to generate music given a set of tags specifying,
e.g., genre\index{Genre} or instruments\index{Instrument}.

\subsubsection{Example 2: Anticipation-RNN Bach Melody Generation}
\label{section:systems:anticipation:rnn}

Hadjeres and Nielsen propose a system named Anticipation-RNN\index{Anticipation-RNN} \cite{hadjeres:anticipation:rnn:arxiv:2017}
for generating melodies
with
unary constraints on notes (to enforce a given note at a given time position to have a given value).
The limitation when using a standard note-to-note iterative strategy for generation by a recurrent network
is that enforcing the constraint\index{Constraint} at a certain time step may retrospectively invalidate
the distribution of the previously generated items,
as shown in \cite{pachet:markov:constraints:ijcai:2011}.
The idea is
to condition the recurrent network (RNN) on some information summarizing the set of further (in time) constraints
as a way to anticipate oncoming constraints,
in order to generate notes with a correct distribution.

Therefore, a second RNN architecture\footnote{Both
	are 2-layer LSTMs \cite{hochreiter:lstm:1997}.},
named Constraint-RNN, is used and it functions backward in time
and ts outputs are used as additional inputs of the main RNN (named Token-RNN),
resulting in the architecture
shown at Figure~\ref{figure:anticipation:rnn:architecture},
with:

\begin{figure}
\center{\includegraphics[scale=0.2]{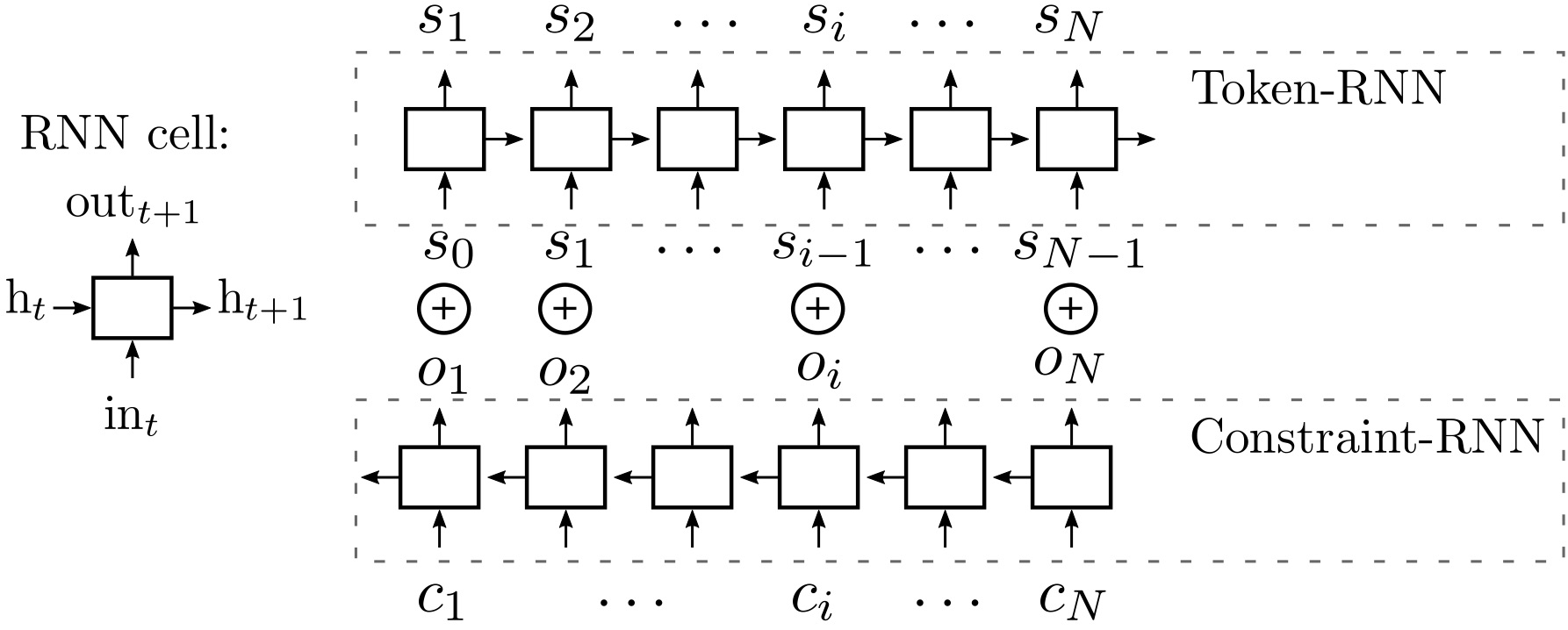}}
\caption{Anticipation-RNN architecture.
Reproduced from \cite{hadjeres:anticipation:rnn:arxiv:2017} with permission of the authors}
\label{figure:anticipation:rnn:architecture}
\end{figure}
\begin{itemize}

\item $c_i$ is a {\em positional constraint\index{Positional constraint}};

\item $o_i$ is the output at index $i$ (after $i$ iterations) of Constraint-RNN
-- it summarizes constraint informations from step $i$ to final step (end of the sequence) $N$.
It will be concatenated ($\oplus$) to input $s_{i-1}$ of Token-RNN in order to predict next item $s_i$.

\end{itemize}



\begin{figure}
\center{\includegraphics[scale=0.25]{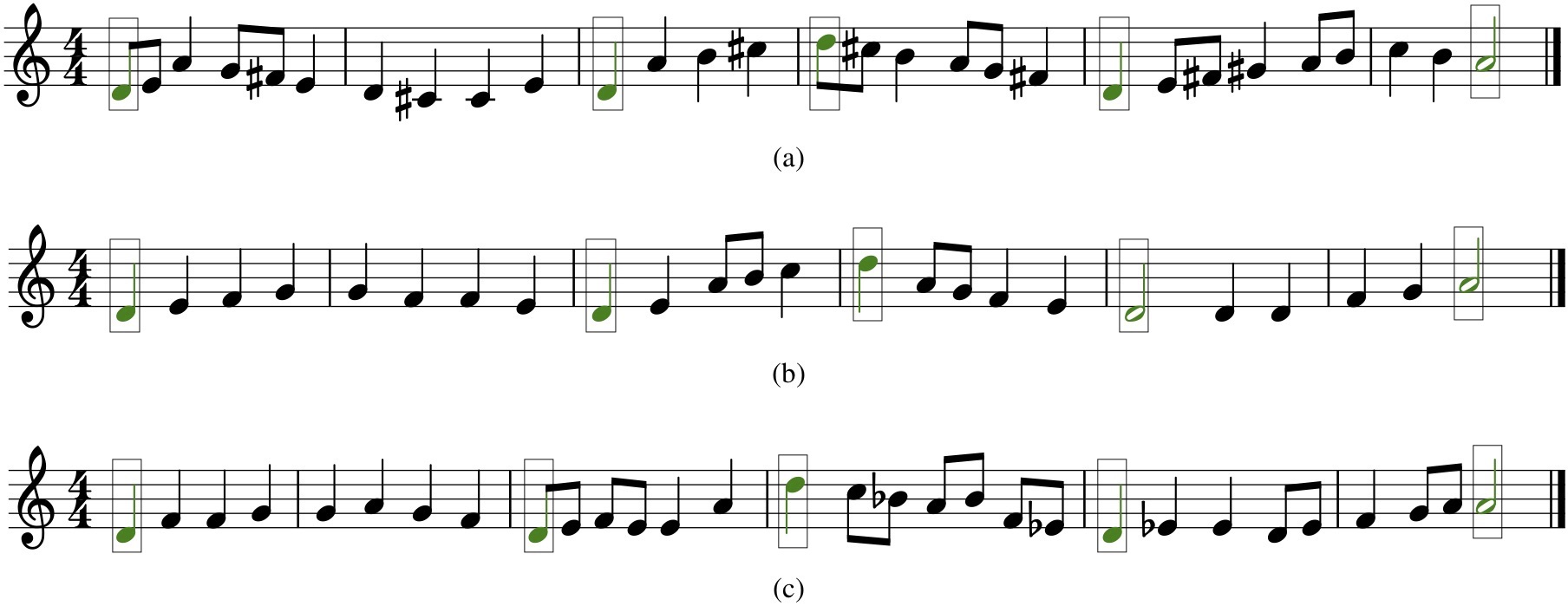}}
\caption{Examples of melodies generated by Anticipation-RNN.
Reproduced from \cite{hadjeres:anticipation:rnn:arxiv:2017} with permission of the authors}
\label{figure:anticipation:rnn:example}
\end{figure}

The architecture has been tested on a corpus of melodies taken from J. S. Bach chorales.
Three examples of melodies generated
with the same set of positional constraints (indicated with notes in green within a rectangle) are shown at Figure~\ref{figure:anticipation:rnn:example}.
The model is indeed able to anticipate each positional constraint by adjusting its direction towards the target (lower-pitched or higher-pitched note).

\subsection{Input Manipulation}
\label{section:control:input:manipulation}

The strategy of {\em input manipulation\index{Input manipulation}} has been pioneered for images\index{Image}
by DeepDream\index{DeepDream} \cite{google:dream:web:2015}.
The idea is that the initial input\index{Input} content, or a brand new (randomly\index{Random} generated) input content,
is incrementally manipulated in order to match a {\em target} {\em property}.
Note that control of the generation is {\em indirect}, as it is not being applied to the output but to the {\em input}, {\em before generation}.
Examples are:

\begin{itemize}

\item {\em maximizing\index{Maximize}} the {\em activation\index{Activation}} of a specific {\em unit\index{Unit}},
to {\em exaggerate} some visual\index{Visual} element\index{Element} specific to this unit,
in DeepDream\index{DeepDream} \cite{google:dream:web:2015};

\item {\em maximizing} the {\em similarity\index{Similarity}} to a given {\em target\index{Target}},
to create a {\em consonant\index{Consonant} melody\index{Melody}},
in DeepHear\index{DeepHear}
\cite{sun:deep:hear};

\item {\em maximizing} {\em both} the {\em content similarity} to some initial image {\em and} the {\em style similarity} to a reference style image,
to perform {\em style transfer}
\cite{gatys:neural:style:2015};

\item {\em maximizing} the {\em similarity} of {\em structure} to some reference music,
to perform {\em style imposition}
\cite{lattner:structure:polyphonic:generation:arxiv:2016}.

\end{itemize}

Interestingly, this is done by reusing
standard training mechanisms\index{Mechanism},
namely back-propagation\index{Back-propagation} to compute the gradients,
as well as gradient descent\index{Gradient descent} to minimize\index{Minimize} the cost\index{Cost}.

\subsubsection{Example 1: DeepHear Ragtime Melody Accompaniment Generation}
\label{section:system:deep:hear:counterpoint}

The DeepHear\index{DeepHear} architecture by
Sun \cite{sun:deep:hear} is aimed at generating ragtime jazz melodies.
The architecture
is a 4-layer stacked autoencoders (that is 4 hierarchically nested autoencoders),
with a decreasing number of hidden units, down to 16 units.

\begin{figure}
\center{\includegraphics[scale=0.16]{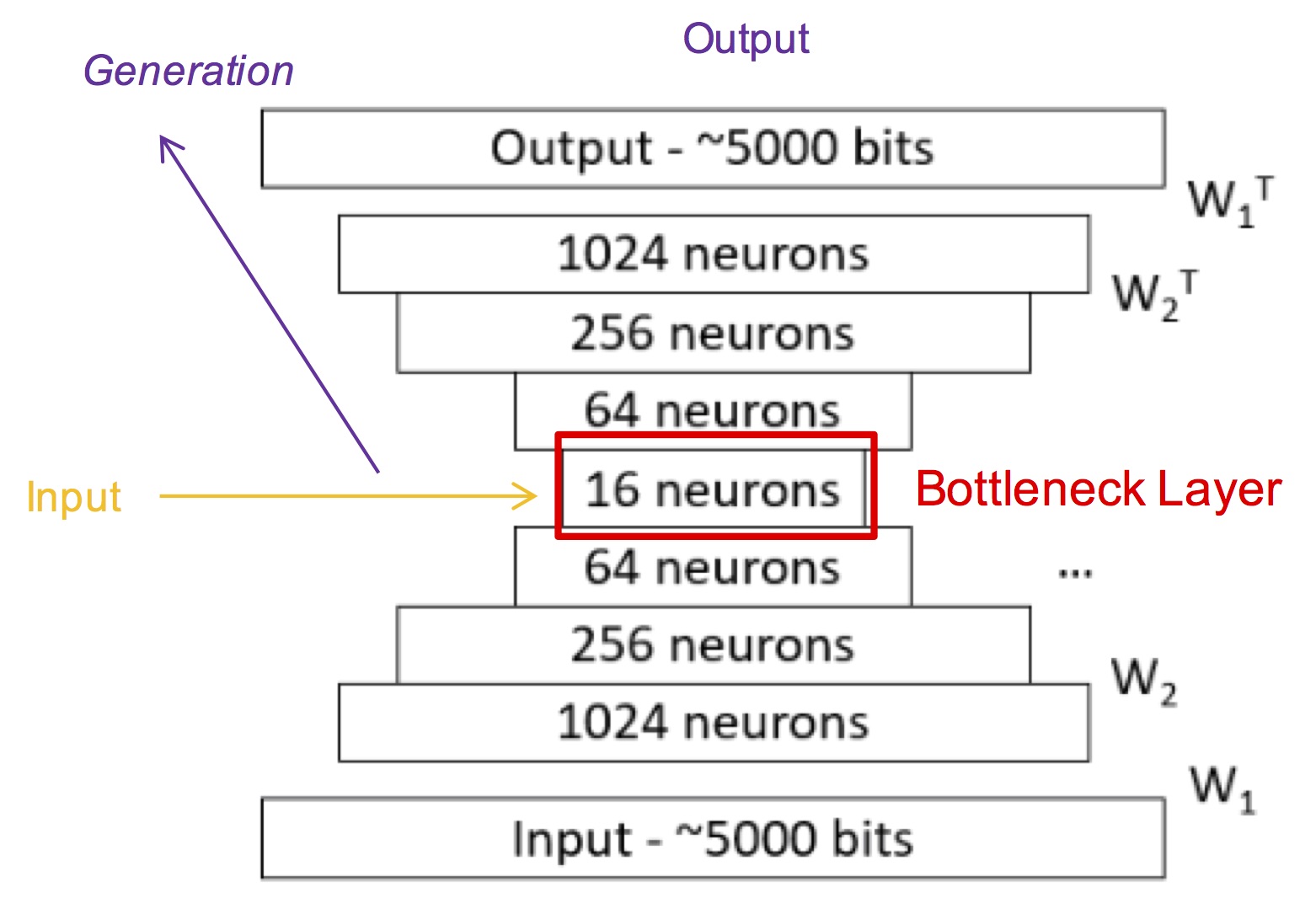}}
\caption{Generation in DeepHear.
Extension of a figure reproduced from \cite{sun:deep:hear} with permission of the author}
\label{figure:hierarchy:autoencoders:generation}
\end{figure}

At first, the model is trained\footnote{Autoencoders are trained with the same data as input and output
	and therefore have to discover significative features in order to be able to reconstruct the compressed data.}
on a corpus of 600 measures of Scott Joplin's ragtime\index{Ragtime} music, split into 4-measure long segments.
Generation is performed by inputing random\index{Random} data as the seed\index{Seed}
into the 16 bottleneck hidden layer units
and then by feedforwarding it into the chain of decoders
to produce an output (in the same 4-measure long format of the training examples),
as shown at Figure~\ref{figure:hierarchy:autoencoders:generation}.

In addition to the generation of new melodies,
DeepHear\index{DeepHear} is used with a different objective:
to harmonize a melody, while using the {\em same} architecture as well as what has already been learnt\footnote{Note that this is a simple example
	of {\em transfer learning\index{Transfer learning}} \cite{goodfellow:deep:learning:book:2016},
	with a same domain and a same training, but for a different task.}.
The idea is to find a label\index{Label} instance of the set of features
i.e. a set of values for the 16 units of the bottleneck hidden layer\index{Bottleneck hidden layer} of the stacked autoencoders
which will result in some decoded output matching as much as possible a given melody.
A simple distance\index{Distance} function is defined to represent the dissimilarity\index{Similarity}
between two melodies (in practice, the number of not matched notes).
Then a gradient descent\index{Gradient!descent} is conducted onto the variables of the embedding,
guided by the gradients corresponding to the distance function
until finding a sufficiently similar decoded melody.
Although this is not a real counterpoint\index{Counterpoint}
but rather the generation of a similar\index{Similarity} (consonant\index{Consonant}) melody,
the results
do produce some naive counterpoint with a ragtime\index{Ragtime} flavor.


\subsubsection{Example 2: VRAE Video Game Melody Generation}
\label{section:system:vrae}

\begin{figure}
\center{\includegraphics[scale=0.3]{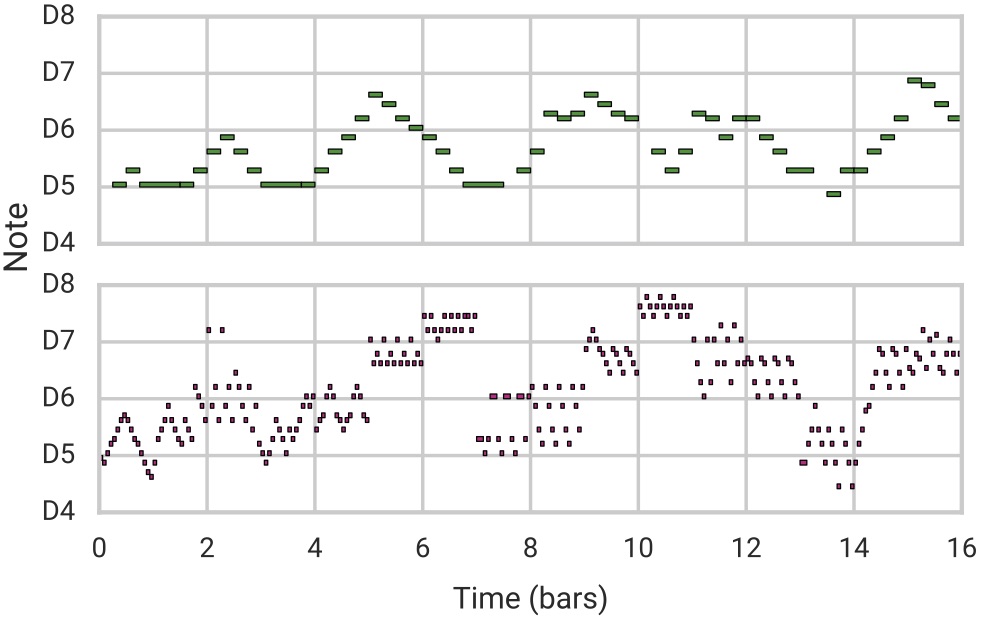}}
\caption{Example of melody generated (bottom) by MusicVAE
by adding a ``high note density'' attribute vector to the latent space of an existing melody (top).
Reproduced from \cite{roberts:hierarchical:latent:arxiv:2018} with permission of the authors}
\label{figure:music:vae:example:density}
\end{figure}

Note that input manipulation
of the hidden layer units of an autoencoder (or stacked autoencoders)
bears
some analogy with variational autoencoders\index{Variational!autoencoder}\footnote{A variational autoencoder\index{Variational!autoencoder}
	(VAE\index{VAE}) \cite{kingma:vae:arxiv:2014}
	is an autoencoder with the added constraint that the encoded representation
	(its latent variables\index{Latent!variable})
	follows some prior probability distribution (usually a Gaussian distribution).
	Therefore, a variational autoencoder is able to learn a ``smooth'' latent space mapping to realistic examples.},
such as for instance
the VRAE\index{VRAE} (Variational Recurrent Auto-Encoder) architecture of Fabius and van Amersfoort \cite{fabius:vrae:arxiv:2015}.
Indeed in both cases, there is some exploration of possible values for the hidden units
(latent variables) in order to generate variations of musical content by the decoder
(or the chain of decoders).
The important difference is that in the case of variational autoencoders, the exploration of values is {\em user-directed},
although it could be guided by some principle,
for example an interpolation to create a medley of two songs,
or the addition or subtraction of an attribute vector capturing a given characteristic
(e.g., high density of notes as in Figure~\ref{figure:music:vae:example:density}).
In the case of input manipulation, the exploration of values is automatically guided by the gradient following mechanism,
the user having priorly specified a cost function to be minimized or an objective to be maximized.

\subsubsection{Example 3: Image and Audio Style Transfer}
\label{section:system:gatys:style:transfer}

\begin{figure}
\center{\includegraphics[width=\textwidth]{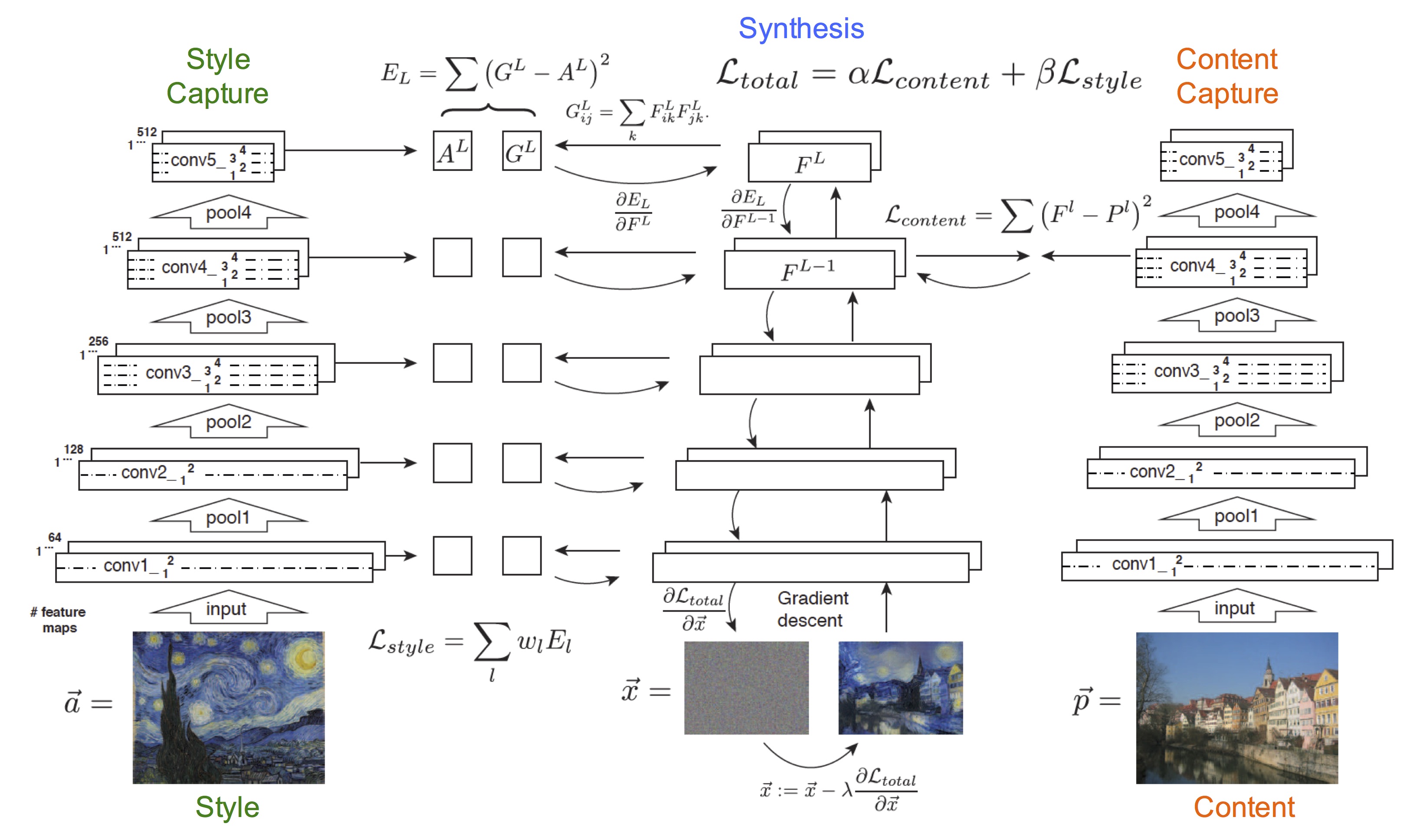}}
\caption{Style transfer full architecture/process.
Reproduced with permission of the authors}
\label{figure:style:transfer:architecture}
\end{figure}

Style transfer has been pioneered by
Gatys {\em et al.} \cite{gatys:neural:style:2015}
for images\index{Image}.
The idea, summarized at Figure~\ref{figure:style:transfer:architecture}, is to use a deep learning architecture to independently capture:

\begin{itemize}

\item the features\index{Feature} of a first image (named the {\em content}),

\item and the {\em style\index{Style}} (the correlations\index{Correlation} between features) of a second image (named the {\em style}),

\item and then, to use gradient\index{Gradient} following to guide the incremental\index{Incremental} modification of an initially random\index{Random} third image,
with the double objective of {\em matching} {\em both} the {\em content} and the {\em style} descriptions\footnote{Note that one may balance
	between content and style objectives through some $\alpha$ and $\beta$ parameters in the $\mathcal{L}_{total}$ combined loss function
	shown at top of Figure~\ref{figure:style:transfer:architecture}.}.

\end{itemize}

\label{section:control:input:manipulation:musical:style:transfer}

Transposing this style transfer technique to music was a natural direction
%
and it has been experimented independently for audio, e.g., in
\cite{ulyanov:audio:style:transfer:web:2016} and \cite{foote:audio:style:transfer:2016},
both using a spectrogram (and not a direct wave signal) as input.
The result is effective, but not as interesting as in the case of painting style transfer,
being somehow more similar to a sound merging of the style and of the content.
We believe that this is because of the
	{\em anisotropy}\footnote{In the case of an image,
	the correlations between visual elements (pixels) are equivalent whatever the direction
	(horizontal axis, vertical axis, diagonal axis or any arbitrary direction), in other words correlations are {\em isotropic}.
	In the case of a global representation of musical content
	(see, e.g., Figure~\ref{figure:generation:strategies}),
	where the horizontal dimension represents time
	and the vertical dimension represents the notes,
	horizontal correlations represent {\em temporal} correlations and vertical correlations represent {\em harmonic} correlations,
	which have very different nature.}
of global music content representation.

\subsubsection{Example 4: C-RBM Mozart Sonata Generation}
\label{section:systems:c-rbm}

\label{section:control:input:manipulation:sampling}



The C-RBM architecture proposed by 
Lattner {\em et al.} \cite{lattner:structure:polyphonic:generation:arxiv:2016}
uses a restricted Boltzmann machine\index{Restricted Boltzmann machine} (RBM\index{RBM}) to learn the {\em local structure},
seen as the {\em musical texture},
of a corpus\index{Corpus} of musical pieces (in practice, Mozart\index{Mozart} sonatas\index{Sonata}).
The architecture is convolutional (only) on the time\index{Time} dimension,
in order to model temporally invariant\index{Invariant} motives\index{Motif},
but not pitch\index{Pitch} invariant motives
which would break the notion of tonality\index{Tonality}.
The main idea is in imposing by {\em constraints\index{Constraint}} onto the generated piece
some more {\em global\index{Global} structure\index{Structure}}
(form\index{Form}, e.g., AABA, as well as tonality\index{Tonality}),
seen as a {\em structural template\index{Template}} inspired from the reference of an existing musical piece.
This is called {\em structure imposition}\footnote{Note that this also some kind of style transfer \cite{dai:music:style:transfer:arxiv:2018},
	although of a high-level structure and not a low-level timbre as in Section~\ref{section:system:gatys:style:transfer}.},
also coined as {\em templagiarism\index{Templagiarism}}
(short for template plagiarism)
by Hofstadter \cite{hofstadter:staring:emi:eye:virtual:music:2001}.
Generation is done by {\em sampling\index{Sampling}} from the RBM with three types of {\em constraints\index{Constraint}}:

\begin{itemize}

\item {\em Self-similarity}, to specify a {\em global structure\index{Structure}} (e.g., AABA) in the generated music piece.
This is modeled by minimizing the distance\index{Distance}
between the self-similarity\index{Similarity}\index{Self-similarity}
matrices of the reference target and of the intermediate solution;

\item {\em Tonality constraint}, to specify a {\em key\index{Key}} (tonality\index{Tonality}).
To estimate the key in a given temporal\index{Temporal} window\index{Window},
the distribution\index{Distribution} of pitch classes\index{Pitch!class}
is compared with the
key profiles\index{Profile}
of the reference;

\item {\em Meter constraint}, to impose a specific {\em meter\index{Meter}} (also named a {\em time signature\index{Time!signature}},
e.g., 4/4) and its related rhythmic\index{Rhytmic} pattern\index{Pattern}
(e.g.,
accent\index{Accent} on the
third beat\index{Beat}).
The relative occurrence of note onsets
within a measure is constrained to follow that of the reference.

\end{itemize}

\begin{figure}
\center{\includegraphics[width=\textwidth]{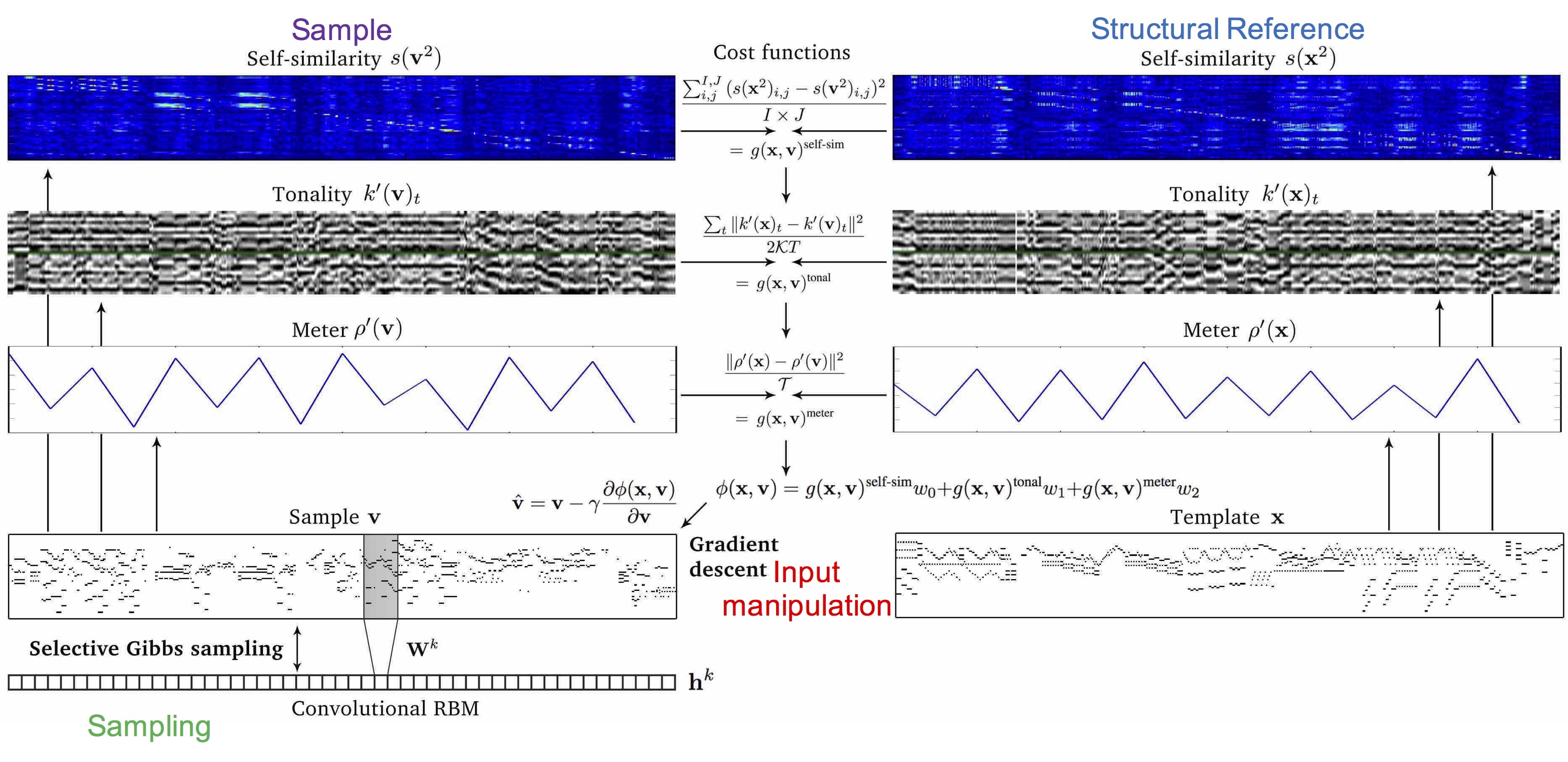}}
\caption{C-RBM Architecture}
\label{figure:crbmc:architecture}
\end{figure}

Generation\index{Generation} is performed via {\em constrained sampling\index{Constrained sampling}},
a mechanism to restrict the set of possible solutions in the sampling\index{Sampling} process
according to some pre-defined constraints\index{Constraint}.
%
The principle of the process (illustrated at Figure~\ref{figure:crbmc:architecture}) is as follows.
At first, a sample is randomly\index{Randomly} initialized\index{Initialize}, following the standard uniform distribution\index{Distribution}.
A step of constrained sampling\index{Constrained sampling}
is composed of
$n$ runs of gradient descent\index{Gradient descent}
to impose the high-level structure,
followed by $p$ runs of {\em selective Gibbs sampling\index{Selective Gibbs sampling}}
to selectively realign the sample onto the learnt distribution.
A {\em simulated annealing\index{Simulated annealing}} algorithm is applied in order to decrease exploration\index{Exploration}
in relation to a decrease of variance\index{Variance} over solutions\index{Solution}.

%

Results are quite convincing.
However, as discussed by the authors,
their approach is not exact, as for instance by the Markov constraints\index{Markov!constraint} approach
proposed in
\cite{pachet:markov:constraints:ijcai:2011}.

\subsection{Reinforcement}
\label{section:control:reinforcement}

The strategy of {\em reinforcement\index{Reinforcement strategy}} is to {\em reformulate} the generation of musical content
as a {\em reinforcement learning problem},
while using the output of a trained recurrent network as an {\em objective}
and adding user defined constraints, e.g., some tonality rules according to music theory, as an {\em additional} {\em objective}.

\label{section:control:reinforcement:learning}
\label{section:architecture:reinforcement:learning}


Let us at first quickly remind the basic concepts of reinforcement learning,
illustrated at Figure~\ref{figure:reinforcement:learning:conceptual:architecture}:

\begin{figure}
\center{\includegraphics[scale=0.25]{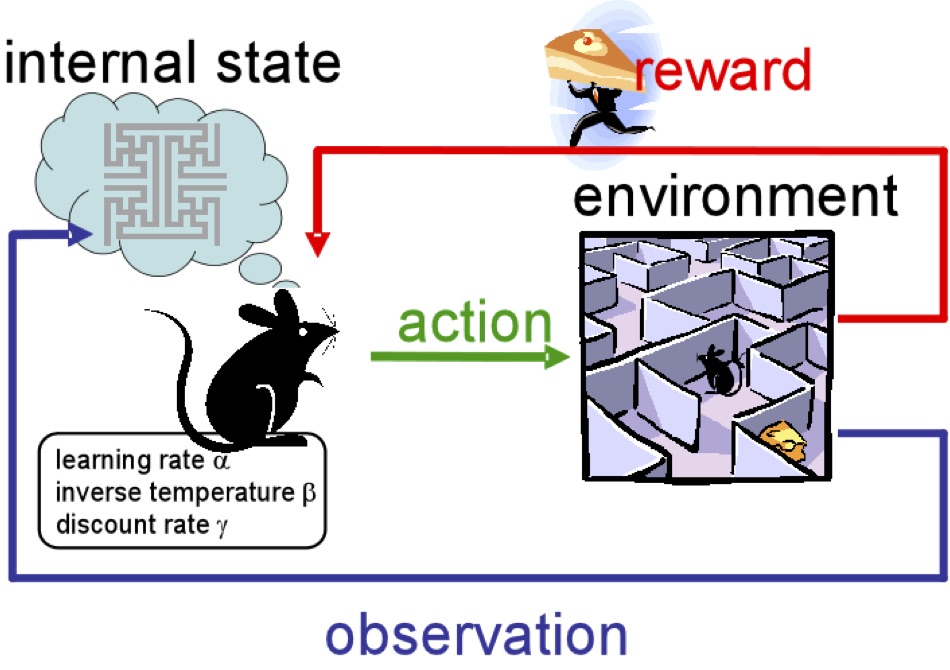}}
\caption{Reinforcement learning (Conceptual model) -- Reproduced from \cite{cyber:rodent:2005}}
\label{figure:reinforcement:learning:conceptual:architecture}
\end{figure}

\begin{itemize}

\item An {\em agent\index{Agent}} sequentially selects and performs {\em actions\index{Action}} within an {\em environment\index{Environment}};

\item Each action performed brings it to a new {\em state\index{State}},

\item with the {\em feedback} (by the environment) of a {\em reward\index{Reward}} ({\em reinforcement signal\index{Reinforcement!signal}}),
which represents some {\em adequation} of the action to the environment (the situation).

\item The objective of {\em reinforcement learning} is for the agent to learn a near optimal {\em policy\index{Policy}} (sequence of actions) in order to maximize its {\em cumulated rewards\index{Cumulated rewards}} (named its {\em gain\index{Gain}}). 

\end{itemize}

Generation of a melody may be formulated as follows (as in Figure~\ref{figure:rl-tuner:architecture}):
the {\em state\index{State}} $s$ represents the musical content (a {\em partial melody}) generated so far
and the {\em action\index{Action}} $a$ represents the selection of next {\em note\index{Note}} to be generated.

\label{section:control:reinforcement:control}


\subsubsection{Example: RL-Tuner Melody Generation}
\label{section:systems:rl-tuner}

\begin{figure}
\center{\includegraphics[scale=0.35]{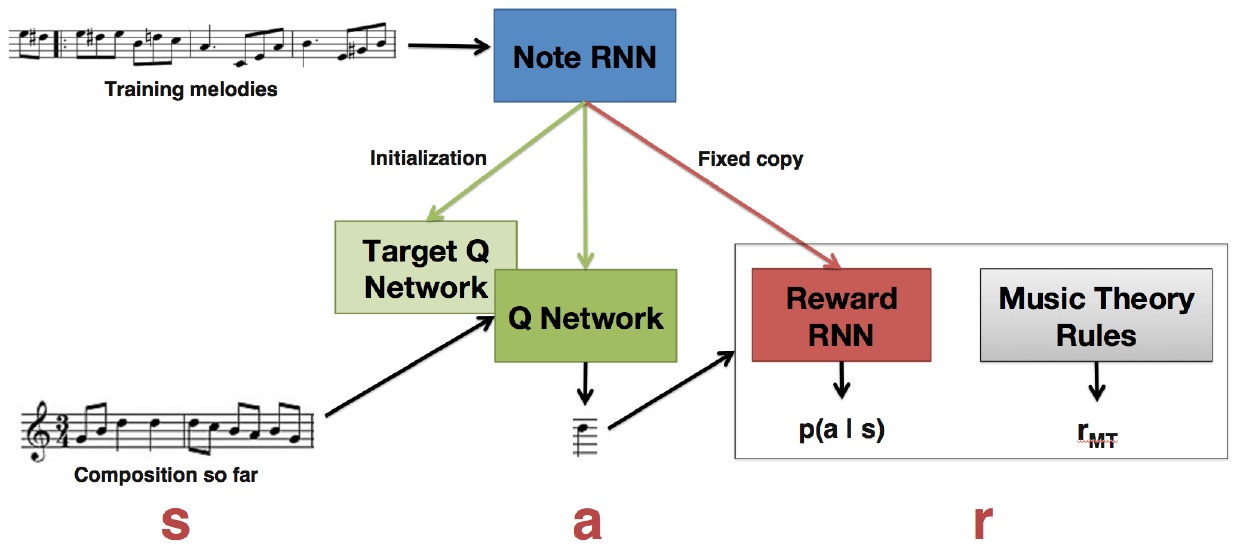}}
\caption{RL-Tuner architecture}
\label{figure:rl-tuner:architecture}
\end{figure}

The {\em reinforcement strategy} has been pioneered by the {\em RL-Tuner\index{RL-Tuner}} architecture
by
Jaques {\em et al.} \cite{jaques:rl:tuner:arxiv:2016}.
The architecture, illustrated at Figure~\ref{figure:rl-tuner:architecture},
consists in two reinforcement learning architectures, named Q Network and Target Q Network\footnote{They use a deep learning implementation
	of the Q-learning algorithm.
	Q Network is trained in parallel to Target Q Network
	which estimates\index{Estimation} the value of the gain\index{Gain})
	\cite{double:q:learning:arxiv:2015}.}
and two {\em recurrent network\index{Recurrent network}}
(RNN\index{RNN}) architectures, named Note RNN and Reward RNN.

After training Note RNN on the corpus,
a fixed copy named Reward RNN is used as a {\em reference} for the reinforcement learning architecture.
The {\em reward} $r$ of Q Network is defined as a combination of two objectives:

\begin{itemize}

\item Adherence to {\em what has been learnt}, by measuring the similarity of the action selected
(next note to be generated)
to the note predicted by Reward RNN in a similar state (partial melody generated so far);

\item Adherence to {\em user-defined constraints} (e.g., consistency with current tonality, avoidance of excessive repetitions\ldots),
by measuring how well they are fulfilled.

\end{itemize}

Although preliminary, results are convincing.
Note that this strategy has the potential for adaptive generation by incorporating feedback from the user.

\subsection{Unit Selection}
\label{section:control:unit:selection}

The {\em unit selection} strategy\index{Unit selection strategy} relies in querying successive {\em musical units\index{Musical unit}}
(e.g., a melody\index{Melody} within a measure\index{Measure})
from a data base and in {\em concatenating\index{Concatenate}} them in order to generate some sequence\index{Sequence} according to some user characteristics.

\subsubsection{Example: Unit Selection and Concatenation Melody Generation}
\label{section:system:brentan:unit:selection}

This strategy has been pioneered by
Bretan {\em et al.} \cite{bretan:unit:selection:arxiv:2016}
and is actually inspired by a technique commonly used in text-to-speech\index{Text-to-speech} (TTS\index{TTS}) systems
and adapted in order to generate melodies (the corpus used is diverse and includes jazz, folk and rock).
The key process here is {\em unit selection\index{Selection}}
(in general each unit is one measure long),
based on two criteria:
{\em semantic\index{Semantic} relevance\index{Relevance}} and {\em concatenation cost\index{Cost}}.
The architecture includes one {\em autoencoder\index{Autoencoder}} and two LSTM\index{LSTM} {\em recurrent networks\index{Recurrent network}}.

The first preparation phase is feature extraction\index{Feature!extraction} of musical units.
10 manually handcrafted\index{Handcrafted} features are considered,
following a {\em bag-of-words\index{Bag-of-words}} (BOW\index{BOW}) approach
(e.g., counts of a certain pitch class\index{Pitch!class}, counts of a certain pitch class rhythm\index{Rhythm} tuple\index{Tuple},
if first note is tied\index{Tied} to previous measure, etc.),
resulting in 9,675 actual features.



The key of the generation
is the process of selection of a best (or at least, very good) successor candidate to a given musical unit.
Two criteria\index{Criteria} are considered:

\begin{itemize}

\item {\em Successor semantic relevance\index{Semantic relevance}} -- It is based on a model\index{Model} of transition\index{Transition} between units,
as learnt by a LSTM\index{LSTM} recurrent network\index{Recurrent network}.
In other words, that relevance\index{Relevance} is based on the distance\index{Distance} to the (ideal) next unit as predicted by the model;

\item {\em Concatenation\index{Concatenation} cost\index{Cost}} -- It is based on another model of transition\footnote{At a more fine-grained\index{Fine-grained} level\index{Level},
	note-to-note\index{Note-to-note} level,
	than the previous one.},
this time between the last note of the unit and the first note of the next unit, as learnt by another LSTM recurrent network.

\end{itemize}

\begin{figure}
\center{\includegraphics[scale=0.35]{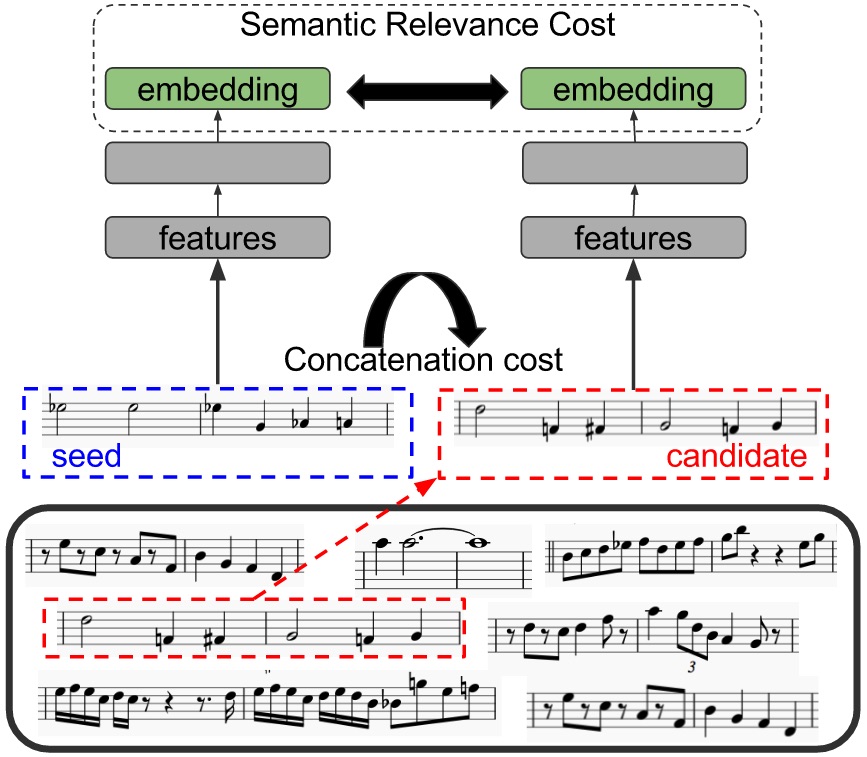}}
\caption{Unit selection based on semantic cost}
\label{figure:unit:selection:semantic-cost}
\end{figure}

The combination of the two criteria (illustrated at Figure~\ref{figure:unit:selection:semantic-cost}) is
handled by a heuristic-based\index{Heuristic} dynamic ranking\index{Ranking} process\index{Process}.
%
%
%
%
%
%
As for a recurrent network, generation is iterated\index{Iterate} in order to create,
unit by unit (measure by measure), an arbitrary\index{Arbitrary} length\index{Length} melody\index{Melody}.

Note that the unit selection strategy
actually provides {\em entry points} for control,
as one may extend the selection framework
based on two criteria: successor semantic relevance and concatenation cost
with user defined constraints/criteria.

\section{Structure}
\label{section:structure}

Another challenge is that most existing systems have a tendency to generate music with
``no sense of direction\index{Sense of direction}''.
In other words, although the style of the generated music corresponds to the corpus learnt,
the music lacks some {\em structure}
and appears to wander without some higher organization\index{Organization},
as opposed to human composed music which
usually exhibits
some global organization (usually named a {\em form\index{Form}}) and identified components,
such as:

\begin{itemize}

\item Overture, Allegro, Adagio or Finale for classical\index{Classical} music;

\item AABA or AAB in Jazz\index{Jazz};

\item Refrain, Verse or Bridge for songs.

\end{itemize}

Note that there are various possible levels of structure.
For instance, an example of finer grain structure is at the level of melodic patterns\index{Melodic pattern}\index{Pattern} that can be repeated,
often
transposed in order to adapt to a new harmonic structure.

Reinforcement (as used by RL-Tuner\index{RL-Tuner} at Section~\ref{section:systems:rl-tuner})
and structure imposition (as used by C-RBM\index{C-RBM} at Section~\ref{section:systems:c-rbm})
are approaches to enforce
some constraints, possibly high-level, onto the generation.
%
An alternative top-down approach is followed by the unit selection strategy\index{Unit!selection strategy}
(see Section~\ref{section:control:unit:selection}),
by incrementally generating an abstract sequence structure and filling it with musical units,
although the structure is currently flat.
Therefore, a natural direction is to explicitly consider and process different levels (hierarchies) of temporality and of structure.

\subsection{Example: MusicVAE Multivoice Generation}
\label{section:system:music:vae}


Roberts {\em et al.} propose a hierarchical architecture named MusicVAE\index{MusicVAE} \cite{roberts:hierarchical:latent:icml:2018}
%
%
%
%
%
%
%
%
following the principles of a variational autoencoder encapsulating recurrent networks (RNNs, in practice LSTMs) such as VRAE\index{VRAE}
introduced at Section~\ref{section:system:vrae},
with two differences:

\begin{itemize}

\item the encoder is a bidirectional RNN\index{Bidirectional!recurrent neural network};

\item the decoder is a hierarchical\index{Hierarchical} 2-level RNN composed of:

\begin{itemize}

\item a high-level RNN named the Conductor
producing a sequence of embeddings\index{Embedding};

\item a bottom-layer RNN
using each embedding as an initial state\footnote{In order to prioritize the Conductor RNN over the bottom layer RNN,
	its initial state is reinitialized with the decoder generated embedding for each new subsequence.}
and also as an additional input concatenated to its previously generated token
to produce each subsequence.

\end{itemize}

\end{itemize}

\begin{figure}
\center{\includegraphics[width=\textwidth]{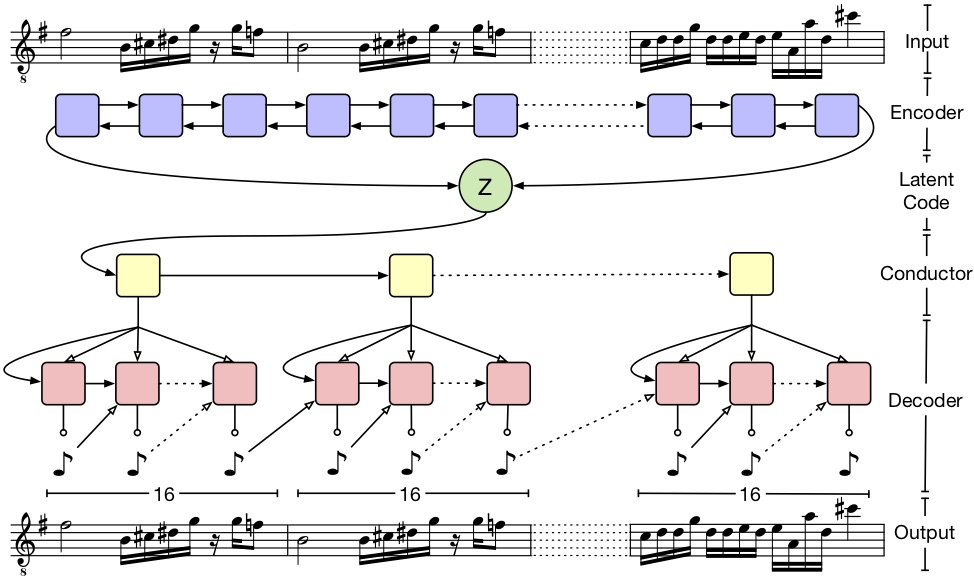}}
\caption{MusicVAE architecture.
Reproduced from \cite{roberts:hierarchical:latent:icml:2018} with permission of the authors}
\label{figure:music:vae:architecture}
\end{figure}

The resulting architecture
is illustrated at Figure~\ref{figure:music:vae:architecture}.
The authors report that an equivalent ``flat'' (without hierarchy)
architecture, although accurate in modeling the style
in the case of 2-measure long examples, turned out inaccurate in the case of 16-measure long examples, with a 27\% error increase
for the autoencoder reconstruction.
Some preliminary evaluation has also been conducted with a comparison by listeners of three versions:
flat architecture, hierarchical architecture
and real music for three types of music: melody, trio and drums,
showing a very significant gain with the hierarchical architecture.

\section{Creativity}
\label{section:creativity}

The issue of the {\em creativity} of the music generated is not only an artistic issue
but also an economic one,
because it raises
a
{\em copyright issue}\footnote{On this issue,
	see a recent paper \cite{deltorn:deep:creations:digital:humanities:2017}.}.

One approach is {\em a posteriori}, by ensuring that the generated music is not too similar\index{Similarity}
(e.g., in not having recopied a significant amount of notes of a melody)
to an existing piece of music.
To this aim,
existing tools to detect similarities in texts may be used.

Another approach, more systematic but
more challenging, is {\em a priori},
by ensuring that the music generated will not recopy\index{Recopy} a given portion of music
from the training corpus\index{Training corpus}\footnote{Note that this addresses the issue
	of avoiding a significant recopy from the training corpus,
	but it does not prevent to {\em reinvent} an existing music outside of the training corpus.}.
A solution for music generation from Markov chains\index{Markov!chain} has been proposed \cite{papadopoulos:avoiding:plagiarism:aaai:2014}.
It is based on a variable order Markov model
and constraints over the order of the generation through some min order and max order constraints,
in order to attain some sweet spot between junk and plagiarism.
However, there is none yet equivalent solution for deep learning architectures.

 
\subsection{Conditioning}
\label{section:originality:conditioning}

\subsubsection{Example: MidiNet Melody Generation}
\label{section:originality:conditioning:midinet}

The MidiNet architecture by Yang {\em et al.} \cite{yang:midinet:ismir:2017}, inspired by WaveNet (see Section~\ref{section:systems:wavenet}),
is based on generative adversarial networks (GAN) \cite{goodfellow:gan:arxiv:2014} (see Section~\ref{section:originality:can}).
It includes a conditioning mechanism incorporating history information (melody as well as chords) from previous measures.
The authors discuss two methods to control\index{Control} creativity\index{Creativity}:

\begin{itemize}

\item by restricting the conditioning\index{Conditioning} by inserting the conditioning data only in the intermediate convolution\index{Convolution} layers of
the generator architecture;

\item by decreasing the values of the two control parameters\index{Control!parameter}
of feature\index{Feature} matching\index{Feature!matching} regularization\index{Regularization},
in order to less enforce\index{Enforce} the distributions\index{Distribution} of real and generated data to be close.

\end{itemize}

These experiments are interesting although the approach remains at the level of some {\em ad hoc} tuning of some hyper-parameters of the architecture.

\subsection{Creative Adversarial Networks}
\label{section:originality:can}
\label{section:originality:can:gan}

Another more systematic and conceptual direction is the concept of creative adversarial networks\index{Creative adversarial networks} (CAN\index{CAN})
proposed by El Gammal {\em et al.} \cite{elgammal:can:arxiv:2017},
as an extension of generative adversarial networks\index{Generative adversarial networks} (GAN\index{GAN}) architecture,
%
%
by Goodfellow {\em et al.}
\cite{goodfellow:gan:arxiv:2014}
which trains simultaneously two networks\index{Neural network}:
%


\begin{itemize}

\item a {\em Generative model} (or {\em generator\index{Generator}}) G,
whose objective is to transform random noise vectors into faked {\em samples\index{Sample}},
which resemble real samples drawn from a distribution of real images; and

\item a {\em Discriminative model} (or {\em discriminator\index{Discriminator}}) D,
that estimates the probability that a sample came from the training data rather than from G.

\end{itemize}

The generator is then able to produce user-appealing synthetic samples (e.g., images or music) from noise vectors.
The discriminator may then be discarded.

\label{section:originality:can:gan:can}


\label{section:experiment:can}

\begin{figure}
\center{\includegraphics[width=\textwidth]{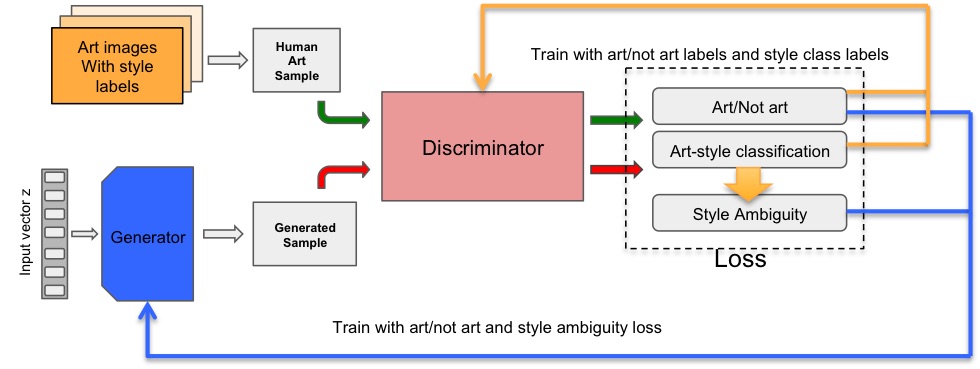}}
\caption{Creative adversarial networks (CAN) architecture}
\label{figure:can:architecture}
\end{figure}

Elgammal {\em et al.} propose in \cite{elgammal:can:arxiv:2017}
to extend a
GAN\index{GAN} architecture
into a {\em creative adversarial networks\index{Creative adversarial networks}} (CAN\index{CAN}) architecture,
shown at Figure~\ref{figure:can:architecture},
%
where the generator\index{Generator} receives from the discriminator\index{Discriminator} not just one but {\em two signals}:

\begin{itemize}

\item the first signal, analog to the case of the standard GAN,
specifies how the discriminator believes that the generated item comes from the training dataset of real art pieces;

\item the second signal is about how easily the discriminator can classify the generated item into {\em established styles}.
If there is some strong ambiguity (i.e., the various classes are equiprobable\index{Equiprobable}),
this means that the generated item is difficult to fit within the existing art styles.

\end{itemize}

These two signals are thus contradictory forces and push the generator to explore the space for generating items
that are at the same time close to the distribution of existing art pieces and with some style originality.
%
%
%
Note that this approach
assumes the existence of a prior style classification
and it also reduces the idea of creativity to exploring new styles
(which indeed has some grounding in the art history).

\section{Interactivity}
\label{section:interactivity}

In most of existing systems, the generation is automated,
with little or no {\em interactivity}.
As a result, local modification and regeneration of a musical content is usually not supported,
the only available option being a whole regeneration
(and the loss of previous attempt).
This is in contrast to the way a musician works, with successive partial refinement and adaptation of a composition\footnote{An example
	of interactive composition environment is FlowComposer \cite{papadopoulos:flow:composer:cp:2016}.
	It is based on various techniques such as Markov models, constraint solving and rules.}.
Therefore, some requisites for interactivity are the incrementality and the locality of the generation,
i.e. the way the variables of the musical content are instantiated.

\label{section:incrementality}
\subsection{Instantiation Strategies}
\label{section:incrementality:strategies}

\begin{figure}
\center{\includegraphics[width=\textwidth]{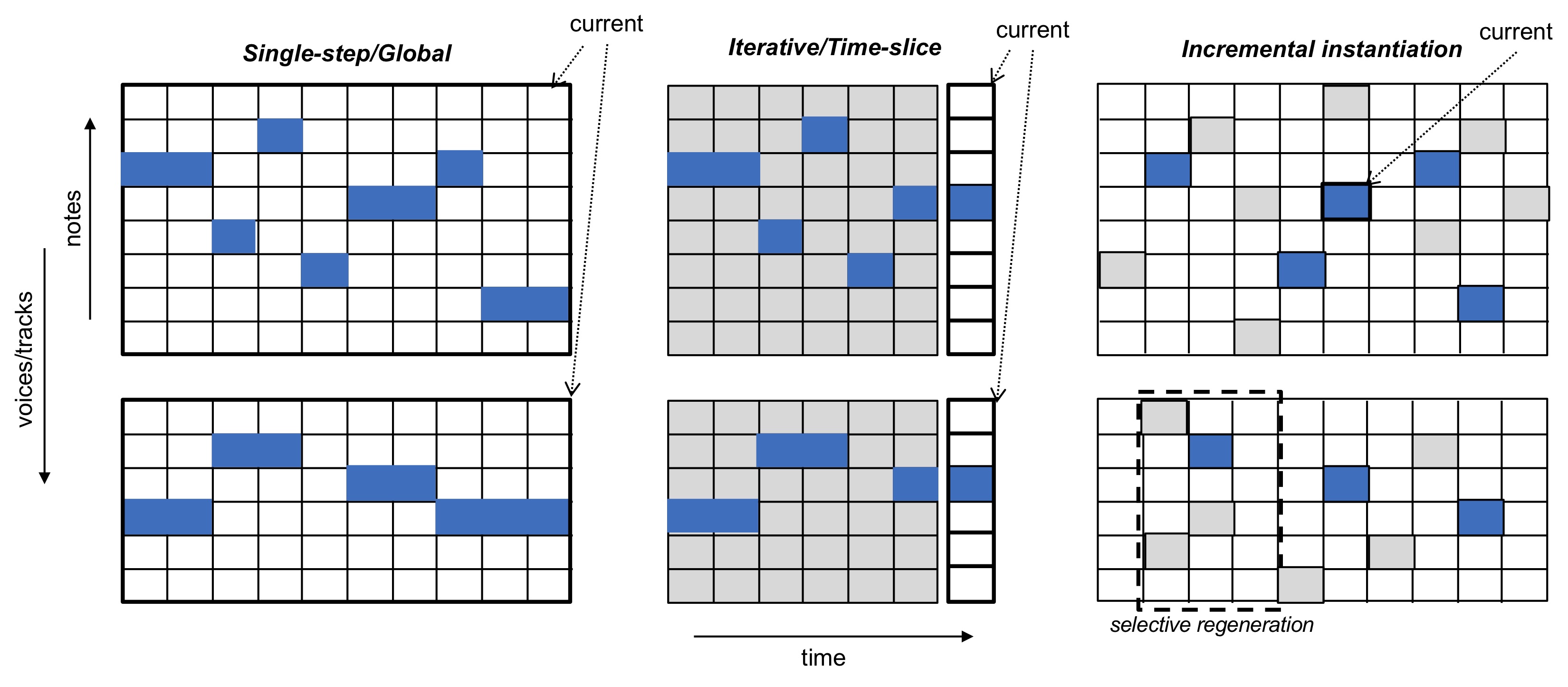}}
\caption{Strategies for instantiating notes during generation}
\label{figure:generation:strategies}
\end{figure}

Let us consider the example of the generation of a melody.
The two
most common strategies
(illustrated at Figure~\ref{figure:generation:strategies})\footnote{The representation shown
	is of type piano roll with two simultaneous voices (tracks).
	Parts already processed are in light grey;
	parts being currently processed have a thick line and are pointed as ``current'';
	notes to be played are in blue.}
for instantiating the notes of the melody
are:

\begin{itemize}

\item
{\em Single-step/Global} --
A global
representation
including all time steps
is generated in a single step by a feedforward architecture.
An example is DeepHear \cite{sun:deep:hear} at Section~\ref{section:system:deep:hear:counterpoint}.


\item
{\em Iterative/Time-slice} --
A time slice
representation
corresponding to
a single time step
is iteratively generated by a recurrent architecture (RNN).
An example is Anticipation-RNN \cite{hadjeres:anticipation:rnn:arxiv:2017} at Section~\ref{section:systems:anticipation:rnn}.


\end{itemize}

Let us now consider an
alternative strategy, {\em incremental variable instantiation}.
It relies on
a global
representation including all time steps.
But, as opposed to single-step/global generation,
generation is
done
{\em incrementally} by progressively instantiating and refining values of variables (notes),
in a non deterministic order.
Thus, it is possible to generate or to {\em regenerate} only an {\em arbitrary part} of the
musical content,
for a specific {\em time interval}
and/or for a specific {\em subset of voices} (shown as selective regeneration in Figure~\ref{figure:generation:strategies}),
without
regenerating the whole content.

\subsection{Example: DeepBach Chorale Generation}
\label{section:experiment:deep:bach}

\begin{figure}
\center{\includegraphics[scale=0.2]{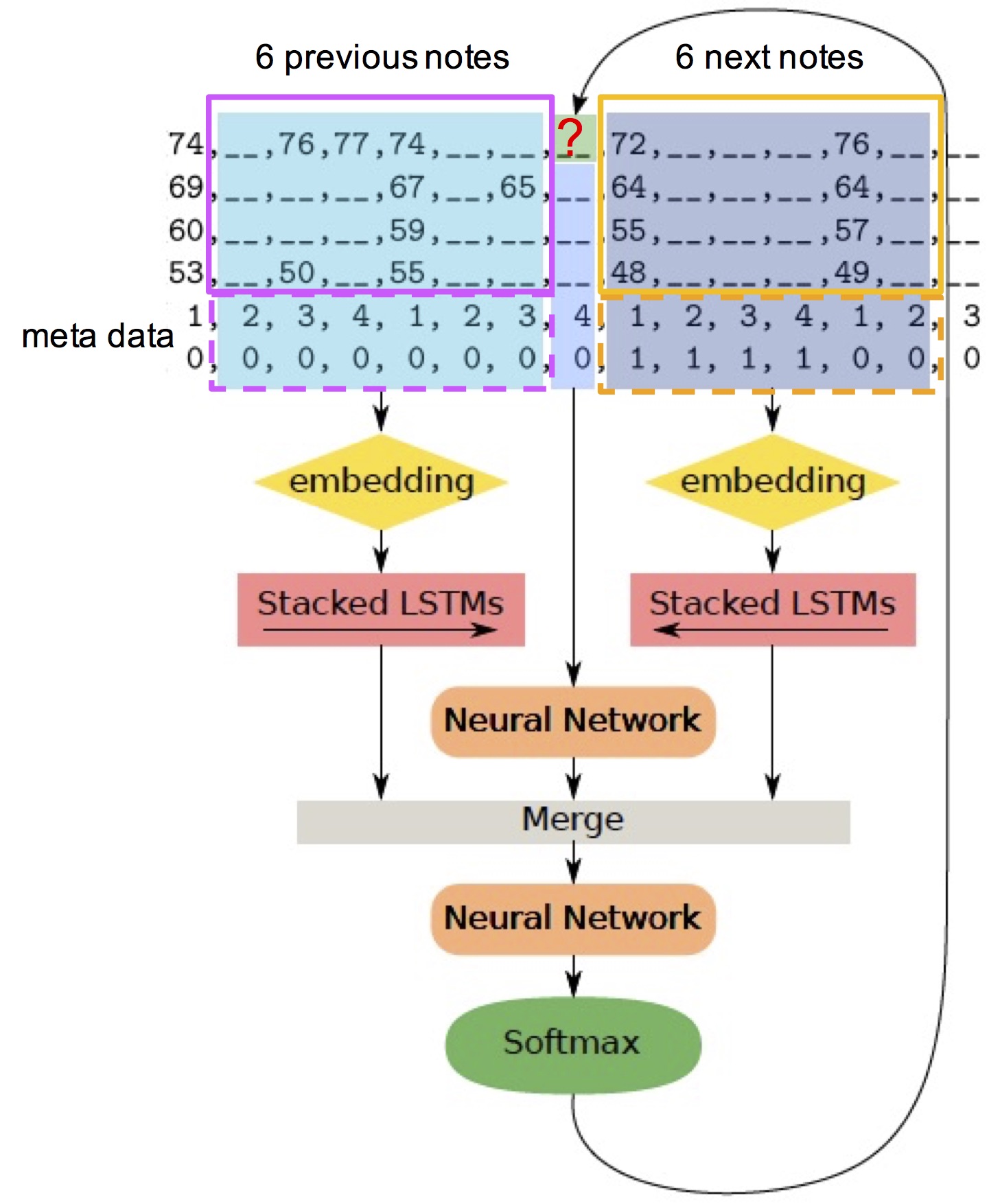}}
\caption{DeepBach architecture}
\label{figure:deep:bach:architecture:sampling}
\end{figure}

This incremental instantiation strategy has been used by Hadjeres {\em et al.} in the DeepBach\index{DeepBach}
architecture
\cite{hadjeres:deep:bach:arxiv:2017}
for generation of Bach\index{Bach} chorales\index{Chorale}\footnote{J. S. Bach chose various given melodies for a soprano
	and composed the three additional ones (for alto, tenor and bass\index{Bass})
	in a {\em counterpoint\index{Counterpoint}} manner.}.
The architecture, shown at
Figure~\ref{figure:deep:bach:architecture:sampling},
combines two recurrent
and two feedforward networks\index{Feedforward network}.
As opposed to standard use of recurrent networks, where a single time direction is considered,
DeepBach architecture considers the two directions {\em forward\index{Forward}} in time and {\em backwards\index{Backwards}} in time.
Therefore, two recurrent networks (more precisely, LSTM)
are used,
one summing up past information and another summing up information coming from the future,
together with a non recurrent network for notes occurring at the same time.
Their three outputs are merged and passed as the input of a final feedforward neural network.
The first 4 lines of the example data on top of the Figure~\ref{figure:deep:bach:architecture:sampling}
correspond to the 4 voices\index{Voice}\footnote{The two bottom lines
	correspond to metadata\index{Metadata} (fermata\index{Fermata} and beat\index{Beat} information), not detailed here.}.
Actually this architecture is replicated\index{Replicate} 4 times, one for each voice\index{Voice} (4 in a chorale).

Training, as well as generation, is not done in the conventional way for neural networks.
The objective is to predict the value of current note for a a given voice
(shown
with a red
?
on top center of
Figure~\ref{figure:deep:bach:architecture:sampling}),
using as information surrounding contextual notes.
The training set is formed on-line by repeatedly randomly selecting a note in a voice from an example of the corpus
and its surrounding context.
Generation\index{Generation} is done
by sampling\index{Sampling},
using a pseudo-Gibbs sampling\index{Pseudo-Gibbs sampling} incremental and iterative algorithm
(shown in Figure~\ref{algorithm:sampling:deep:bach}, see details in \cite{hadjeres:deep:bach:arxiv:2017})
to produce a set of values (each note) of a polyphony, following the distribution that the network has learnt. 

\begin{figure}
Create four lists $V = (V_1; V_2; V_3; V_4)$ of length $L$;\\
Initialize them with random notes drawn from the ranges of the corresponding voices\\
{\bf for} $m$ from $1$ to $max\,number\,of\,iterations$ {\bf do}\\
Choose voice $i$ uniformly between $1$ and $4$;\\
Choose time $t$ uniformly between $1$ and $L$;\\
Re-sample $V_i^t$ from $p_i(V_i^t | V_{\symbol{92} i, t}, \theta_i)$\\
{\bf end for}
\caption{DeepBach incremental generation/sampling algorithm}
\label{algorithm:sampling:deep:bach}
\end{figure}

The advantage of this method is that generation may be tailored.
For example, if the user changes only one or two measures of the soprano voice,
he can resample only the corresponding counterpoint voices for these measures.

\label{section:interactivity:deep:bach}

\begin{figure}
\center{\includegraphics[scale=0.1]{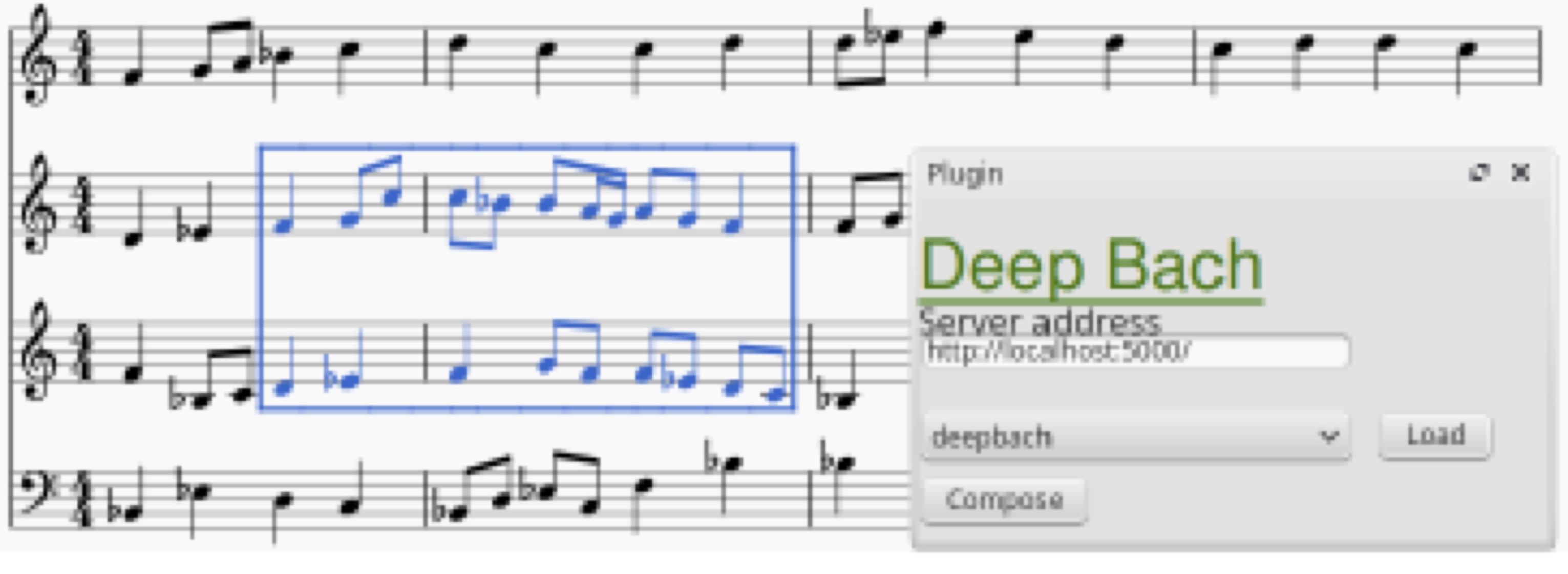}}
\caption{DeepBach user interface}
\label{figure:deep:bach:user:interface}
\end{figure}

The user interface of DeepBach,
shown at Figure~\ref{figure:deep:bach:user:interface},
allows the user to interactively select and control global or partial (re)generation of chorales.
It opens up new ways of composing Bach-like chorales for non experts
in an interactive manner,
similarly to what is proposed by FlowComposer for lead sheets
\cite{papadopoulos:flow:composer:cp:2016}.
It is implemented as a plugin for the MuseScore music editor.
%

\section{Conclusion}
\label{section:conclusion}

The use of deep learning architectures and techniques for the generation of music (as well as other artistic content)
is a growing area of research.
However, there remain
open challenges such as control, structure, creativity and interactivity,
that standard
techniques do not directly address.
In this article, we have discussed a list of challenges,
introduced some strategies to address them
and have illustrated them through examples of actual architectures\footnote{A more complete survey
	and analysis is \cite{briot:dlt4mg:springer:2018}.}.
We hope that the analysis presented in this article will help at a better understanding of issues and possible solutions
and therefore may contribute to the general research agenda of deep learning-based music generation.

\subsubsection*{Acknowledgements}
We thank Ga\"etan Hadjeres and Pierre Roy for related discussions.
This research was partly conducted within the Flow Machines project which received funding from the European Research Council
under the European Union Seventh Framework Programme (FP/2007-2013) / ERC Grant Agreement n. 291156.

\bibliographystyle{alpha}      
\bibliography{dl4mg}   

\end{document}